# The visualisation of two-dimensional dose distribution on X-ray CT using radiochromic film

Nobuyoshi Tanki[1], Toshizo Katsuda[2], Masashi Sasaki[3], Rumi Gotanda[4], Tatsuhiro Gotanda[4], Shinya Imai[5], Yasuyuki Kawaji[6], Atsushi Noguchi[7]

[1] Brain Activity Imaging Center, ATR-Promotions Inc., Kyoto, Japan
[2] Dept. of Medical Radiation Science, Shizuoka College of Medicalcare Science, Shizuoka, Japan
[3] Faculty of Health Science, Butsuryo College of Osaka, Osaka, Japan
[4] Dept.of Radiological Technology, Faculty of Health Science and Technology, Kawasaki University of Medical Welfare, Okayama, Japan
[5] Dept. of Radiological Sciences, Faculty of Health Science, Morinomiya University of Medical Sciences, Osaka, Japan
[6] Dept. of Radiological Science, Faculty of Health Sciences, Junshin Gakuen University, Fukuoka, Japan
[7] Seishokai Aoi Hospital, Hyogo, Japan

*Abstract*— **X-ray CT dose measurement has mainly been performed using an ionization chamber dosimeter. Therefore, the dose distribution has not been sufficiently studied. We investigated the importance of evaluating the two or three-dimensional dose distributions for X-ray computed tomography (CT). To confirm this purpose, we investigated the effects of phantom size and exposure parameters on the phantom diameter.**

**We performed 12 scans using XR-QA2 film and cylindrical acrylic phantoms of two lengths. The tube current and slice thicknesses were varied as exposure parameters. The dose distribution of the primary and scattered radiation was visualised using ImageJ. The results were evaluated using the profile curves, three-dimensional surface plots, and subtraction dose images.**

**We observed changes in the dose distributions for scans with and without phantoms. However, no significant difference in the dose distribution was observed with changes in the lengths of the phantom. To visualise the dose distribution, three-dimensional surface plots and subtraction images were found helpful. We must confirm that the dose distribution does not change for phantoms with multiple diameters in future studies. For using a clinical application, accurate quantitative assessment of dose distribution requires improved accuracy of the calibration curve.**

## I. Introduction

Recently, X-ray computed tomography (CT) examinations have become widespread in medicine.[1] Quality control (QC) of the exposure dose is necessary to establish good and stable performance of the X-ray CT system.[2,3] Without an accurate radiation dose, it is difficult to perform appropriate radiation examinations of patients. The routine dose assessment of the X-ray CT is characterised by the CT dose index (CTDI), which is measured using an ionisation chamber;[4] however, not all hospitals have ionisation chambers. In addition, while dosimetry using an ionisation chamber is standard for estimating CTDI, it becomes time-consuming when there are many measurement points. Radiochromic film (RCF) is a useful alternative tool for measuring radiation exposure.[5,6] RCF dosimetry can be performed using the RCF with a commercial scanner, which reduces system installation costs. Furthermore, the ionisation chamber can only measure radiation doses in a selected area where the dosimeters are placed; therefore, CTDI means are averaged measurements over limited areas.

Here, we attempt to establish a methodology for three-dimensional (3D) RCF dosimetry to obtain an accurate 3D dose distribution in X-ray CT. For this method, detailed information regarding the two-dimensional (2D) dose distribution is required.[7] Various factors relating to the CT equipment can cause changes in the dose distribution.[8,9] In this study, we used 10-cm-diameter cylindrical phantoms, used to simulate the head of an infant,[10] of two lengths. The two phantom lengths were selected to investigate changes in the dose distribution with length, as suggested in a previous study.[11] Mori et al. investigated the relationship between the phantom length and cone beam CT dose using a silicon photodiode. We have considered the effects of phantom length on the dose distribution contributed by the scattered radiation. Tomic et al. [12] reported changes in beam quality within CTDI phantoms using Monte Carlo simulation and RCF methods. The advantage of the RCF method over other dosimetry methods is that it provides a dose distribution; however, many previous studies of X-ray CT dosimetry have used profile curves instead of dose distributions.[13-15]

To apply RCF dosimetry to X-ray CT for QC, the characteristics of the dose distribution must be confirmed under normal conditions. Measuring the contribution of the scattered radiation and obtaining high-resolution dose distributions over the entire exposure area are difficult when using an ionisation chamber.[16] The radiation exposure produced by the X-ray CT is the summation of the primary and scattered radiations.[17] In CT scans, several images with various slices per patient are obtained, and the dose of each slice from scattered radiation differs from the others.[18] The verification of the scattered radiation effect in the phantom leads to an accurate evaluation of radiation exposure by X-ray CT dosimetry using RCF.

In this study, we confirmed the changes in dose distribution with different object shapes and exposure parameters using RCF and a film-folding phantom. Moreover, we confirmed the visualisation method such as pseudo-colour map or three-dimensional plot as displaying dose distribution. The purpose of this investigation was the confirmation of the importance of evaluating the 2D or 3D dose distribution for dosimetry in diagnostic X-ray CT imaging.

II. MATERIALS AND METHODS

*A. Experimental design*

We performed 12 experiments under four exposure conditions and three types of phantom conditions in this study. We used two acrylic phantoms of different lengths (10 cm and 30 cm) with the same diameter (10 cm) in our experiments, as shown in Fig. 1. For a comparison of the acquired data with these phantoms, the RCF was also exposed without a phantom. To expose the RCF in the X-ray CT system, we used tube currents of 100 and 300 mA and slice thicknesses of 8 and 16 mm. The tube voltage and time per gantry rotation was fixed for all measurements at 120 kV and 1 s, respectively. We used a clinical X-ray CT system (Aquilion Lightning; Canon Medical Systems, Ohtawara, Japan) for X-ray exposure. To simplify the interpretation of the acquired data, we only used a single scan. The starting point of the X-ray exposure could not be fixed owing to the mechanical limitations.

GAFChromic XR-QA2 films (Ashland, Inc., USA) were used in this study. The RCF was cut out 5 mm larger than the size of the phantom, and data from the region outside the edge of the phantom were discarded. The RCF cutting process was performed 24 h before radiation exposure.

*B. Scanning RCF and image processing*

A commercial image scanner (Epson ES-2200, Seiko Epson Co., Nagano, Japan) was used to digitise the exposed RCFs. To remove the Moiré artefact, a protective film with a liquid crystal display (LCD-230W; Sanwa Supply Inc., Okayama, Japan) was placed over the reading face of the scanner. The films were read with a 48-bit, 100-dpi resolution in RGB mode. From the scanned images, we separated three colour channels and used the red channel data owing to its high sensitivity for the lower radiation doses. To measure the increase in the density of the RCFs following exposure and to eliminate the noise due to the yellow polyester layer, the subtraction process was performed on the scanned images before and after exposure.

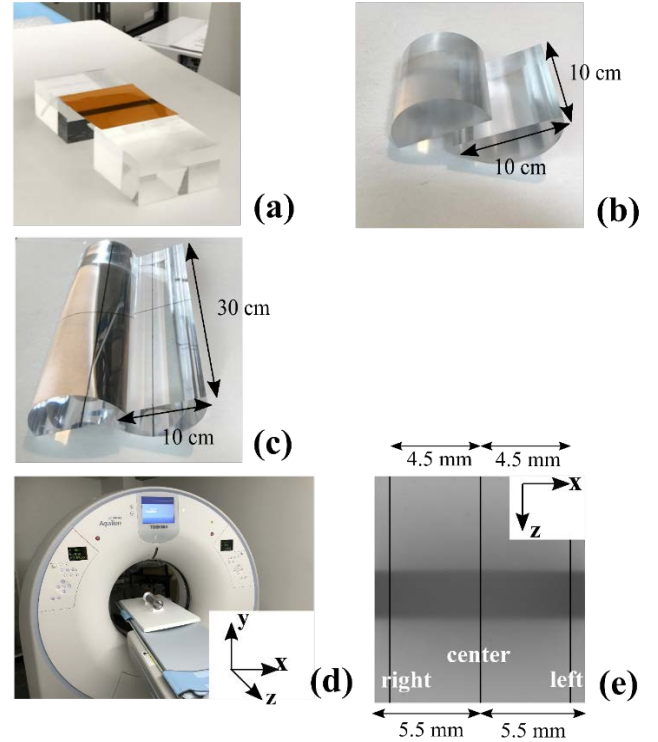

Figure 1 Experiment setup without phantom (a). The semi-cylindrical phantoms of short (b) and long length (c). In this article, the three axes were defined as (d). The phantom was exposed on head-first direction, and 'right' and 'left' are designated based on head-first positioning in this article. The profile curves were taken at five positions shown in (e).

*C. Conversion of pixel value to dose*

A commercial image scanner (Epson ES-2200, Seiko Epson Co., Nagano, Japan) was used to digitise the exposed RCFs. To remove the Moiré artefact, a protective film with a liquid crystal display (LCD-230W; Sanwa Supply Inc., Okayama, Japan) was placed over the reading face of the scanner. The films were read with a 48-bit, 100-dpi resolution in RGB mode. From the scanned images, we separated three colour channels and used the red channel data owing to its high sensitivity for the lower radiation doses. To measure the increase in the density of the RCFs following exposure and to eliminate the noise due to the yellow polyester layer, the subtraction process was performed on the scanned images before and after exposure.

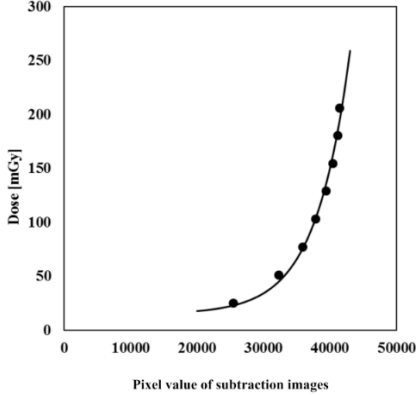

Figure 2 The dose calibration curve at 120 kVp. The measurements were fitted by the following equation: $y = 15 + 5.4677 \times 10^{-2} \cdot e^{1.9542 \times 10^{-4} x}$. The coefficient of determination ($r^2$) was 0.99477.

### D. Scanning RCF and image processing

We measured the colour changes of RCF by radiation exposure to quantitatively evaluate the radiation dose distribution. The absorbed doses were measured by a semiconductor dosimeter (NOMEX multimeter; PTW Inc., Freiburg, Germany) to verify the relationship between the exposure doses and pixel value changes in the RCF. The dose-calibration curve shown in Fig. 2 was calculated using Kaleida Graph 4.0 (HULINKS Inc.; Tokyo, Japan). The data were collected in the range of 25–200 mGy at 120 kVp. The dose-calibration curve was fitted using the following equation: $y = a + b \cdot e^{cx}$

The selection of the fitting function was based on previous studies.19-21 The fitting curve coefficients (dimensionless number) were used in Eq. (1) were as follows: $a = 15$; $b = 5.4677 \times 10^{-2}$; $c = 1.9542 \times 10^{-4}$

### E. Calculation of dose distribution map and visualisation of dose distribution

The data converted to doses were displayed by ImageJ software (version 1.52; National Institutes of Health, Bethesda, MD, USA) as the dose distribution map. The visible dose range on the dose distribution map was adjusted using the setting parameters of the software. The ImageJ software was used for image analysis and for pseudo-colour map creation.

To investigate the effects of the phantom size and exposure parameters, we denoted the three-plane coordinate system as shown in Fig. 1(d). The profile curves (Fig. 1(e)) were obtained along the z-axis direction.

Furthermore, we calculated the subtraction images to investigate the influence of the scattered radiation. The images of the dose distribution map before subtraction were registered by the Align Slice plugin in ImageJ. To calculate the subtraction images for the comparison of different slice thicknesses, the upper edge of the direct exposure area on each image was registered. The dose distributions were displayed by the Interactive 3D Surface Plot plugin in ImageJ, as this plugin facilitates visual evaluation of the differences in the dose distribution.

### III. RESELTS

To quantify the differences in the dose maps, profile curves were taken along the short-axis and long-axis directions of each dose distribution map.

### A. Comparison of dose distribution for different phantom lengths

We studied the dose distributions for the long-length phantom (30 cm) and the short-length phantom (10 cm). To investigate the effects of the dose and dose distribution in and near the exposure area under the influence of scattered radiation, we compared our results with and without the phantom under the same exposure conditions.

The dose distribution near the direct exposure area focused along the x-axis for no-phantom, short-, and long-length phantoms are shown in Fig. 3. The shape of the dose distribution differed for each case. We observed that the doses on the left side were higher without a phantom, whereas the doses at the centre decreased with the phantom. In the presence of phantoms, the doses on the left and right sides of the phantoms were higher. The 3D plot in Fig. 3 allows us to observe a wider range of dose distributions compared to that using the profile curve, which only shows the doses along a certain line. The drawback of this display method is the lack of quantifiable indicators.

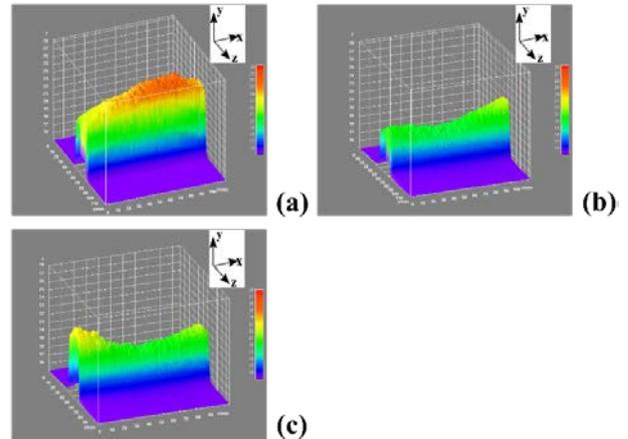

Figure 3 An example of 3D dose distribution with dose value as amplitude without phantom (a), with short phantom (b), and with long phantom (c). The data on the z-axis direction was trimmed in (c).

The dose profile curves along the z-axis at thicknesses of 8 mm and 16 mm are shown in Figs. 4 and 5. These results indicate that the doses without a phantom are higher for each exposure condition, except for Fig. 5 (f), which shows a comparable exposure with the phantom for the 300-mA exposure (right). This observation is consistent with the data shown in Fig. 3 (c). In both figures, the profile curves without the phantom are steep, whereas those with the phantoms are gentle.

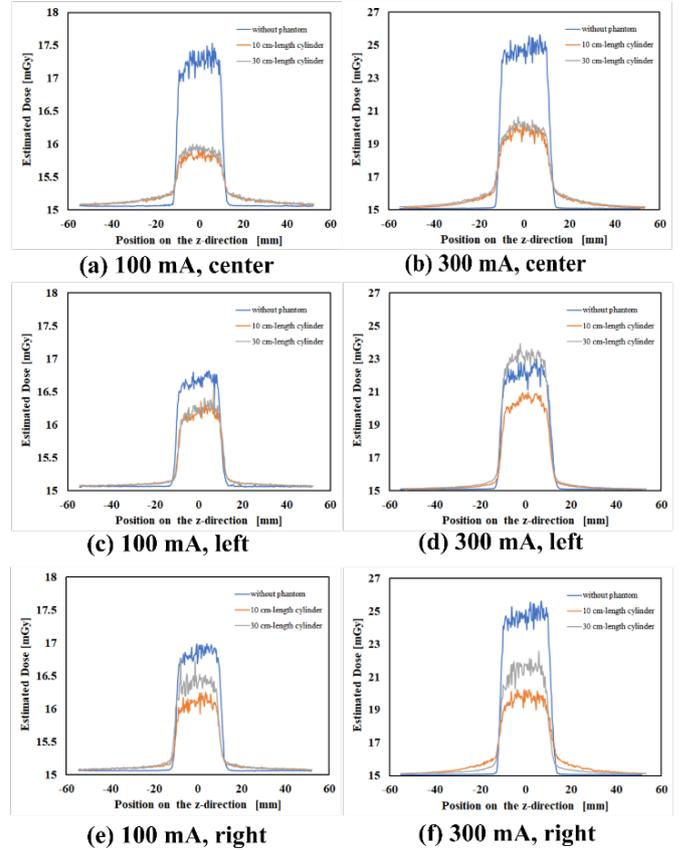

(a) 100 mA, center    (b) 300 mA, center

(c) 100 mA, left    (d) 300 mA, left

(e) 100 mA, right    (f) 300 mA, right

Figure 5 The dose profile curves of 16-mm slice thickness

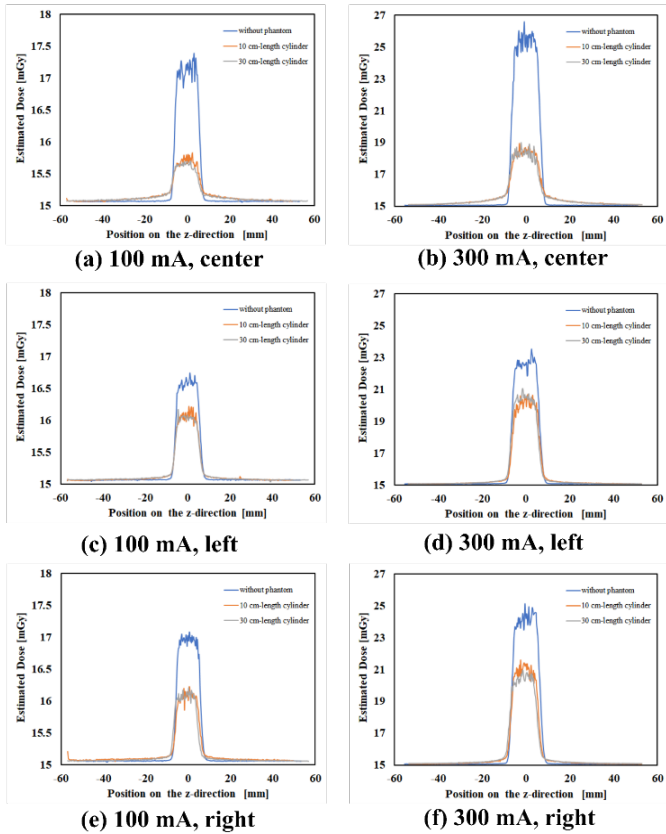

(a) 100 mA, center    (b) 300 mA, center

(c) 100 mA, left    (d) 300 mA, left

(e) 100 mA, right    (f) 300 mA, right

Figure 4 The dose profile curves of 8-mm slice thickness

Figure 6 shows a comparison of the subtraction images with and without phantoms under the same conditions. The low doses observed around the direct exposure area may be due to scattered radiation.

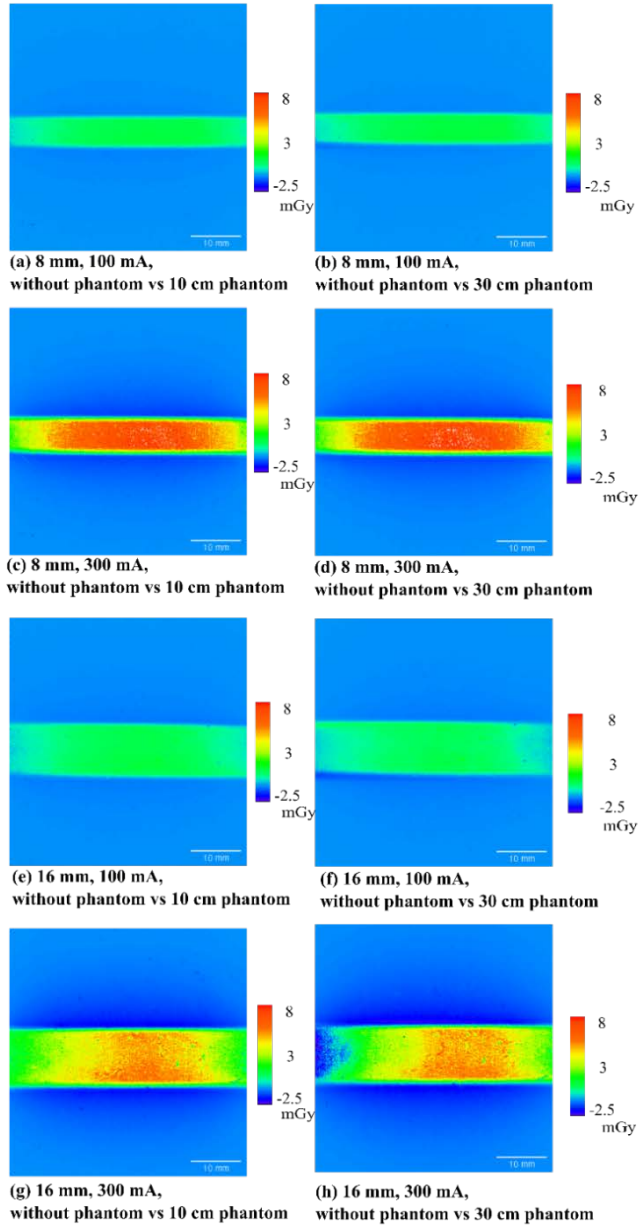

Figure 6 The subtraction images of dose distribution maps

## B. Comparison of dose distribution for different exposure parameters

It was previously reported that the scattered radiation increases with the expansion of the exposed area, and that the dose increases with increases in the tube current;[22] however, the definite effects of scattered radiation have not yet been verified. Here, we demonstrated the effects of tube current (Fig. 7) and slice thickness (Fig. 8) on the subtraction images. The purpose of creating the subtraction images was to investigate the effects of scattered radiation for the selected phantom length. The subtraction images with phantoms displayed low-dose areas around the direct exposure in Fig. 7. Similar behaviour was observed under high-dose conditions, as shown in Fig. 8. To compare the subtraction images shown in Figs. 8 (e) and 8 (g), another subtraction image was calculated (data not shown), indicating a small dose difference (less than 1 mGy) on the lower left side of the image.

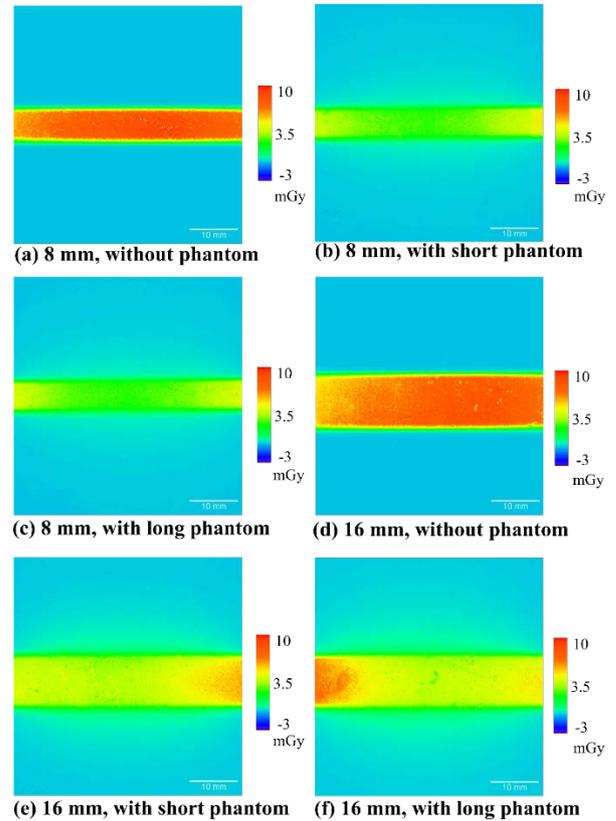

Figure 7 The subtraction images of dose distribution maps with different tube currents. The low-dose images were subtracted from the high-dose images.

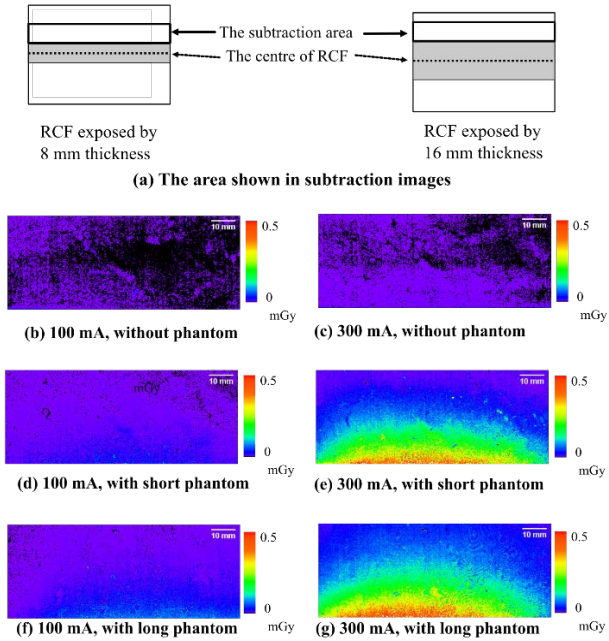

Figure 8 The subtraction images of dose distribution maps between different slice thicknesses. The subtraction images were created by the two images shown in (a). The grey-coloured area shown in (a) indicates the radiation-exposed area.

## IV. DISCUSSION

### A. Comparison of dose distribution for different phantom lengths

The 3D surface plot and profile curves were analysed to investigate the effects of the radiation scattered by the phantom. In Figs. 4 and 5, the difference in the profile curve shapes near the direct exposure areas for the scans with and without phantoms indicates the scattered radiation. In this study, data without the scattered radiation could not be measured. Because the position of the RCF is fixed in the CT system, the effects of the scattered radiation from objects such as the CT bed cannot be excluded. The shapes of the dose distribution curves with phantoms (shown in Fig. 3) are comparable to those of a 32-cm-diameter CTDI phantom obtained using Monte Carlo simulations in a previous study;[23] however, our results differed from those of a 16-cm-diameter phantom investigated in the same study, which was closer to the size to our phantom. Their study suggests that the phantom diameter also affected the measurement dose. The ionisation chamber cannot measure the area outside the dosimeter, whereas the RCF method allows measurements of scattered radiation by adjusting the size of the RCF. The profile curves with and without phantoms (Figs. 4 and 5) display the same tendency as the distribution of the primary radiation and scattered radiation depicted in a previous study.[24]

In contrast, according to Fig. 6, the selected lengths of the phantoms did not seem to affect the dose distribution. We hypothesised that a phantom with sufficient length is necessary for accurate dosimetry, considering the effects of scattered radiation based on previous reports.[11,18] The statistical values for the dose distribution map were acquired to evaluate any differences in the statistical values of the images of the long and short phantoms. Our results indicate that the length of the phantom does not affect the image statistical values. Previous studies on CTDI measurements of 16-cm- or 32-cm-diameter phantoms indicated that a phantom of sufficient length was necessary for accurate dose measurements with scattered radiation; however, the estimated optimum phantom length differed in each study.[11,18,25] In this study, we did not observe any significant differences in the dose distribution with changes in the phantom lengths for our 10-cm-diameter phantoms. We believe that our method may be useful for obtaining accurate doses for various object sizes because it allows measurement of various doses, including those outside the exposure range.

### B. Comparison of dose distribution for different exposure parameters

The profile curve analysis was performed, and the subtraction images were calculated to evaluate the effects of the exposure parameters on the scattered radiation. The scattered radiation increases with the expansion of the exposure area; however, the scattered radiation has not been quantitatively evaluated in the literature. In Fig. 8, the subtraction images near the direct exposure areas displayed small differences in the dose distribution for the long and short phantoms.

The treatment plan for radiation therapy such as intensity modulation radiation therapy is compared with the actual results measured by a dosimeter[26-28] owing to the uncertainty of the planned dose calculated by the treatment planning device and the statistical uncertainty of Monte Carlo simulations.[29] Dose estimation by Monte Carlo simulation has been used for X-ray CT, but it is not always compared with actual results. This is because X-ray CT does not cause immediate health concerns compared to radiotherapy, which utilises a high-energy radiation and may cause serious side effects because of errors in irradiation doses. The absorbed doses with various types of X-ray CTs can differ even under the same exposure conditions.[30] Therefore, it is desirable to perform actual measurements and verification at each facility to determine accurate exposure dose control. Our dosimetry method may be useful for such purposes. The first goal of our study was to establish a single-scan 3D dosimetry method for X-ray CT systems with RCF and film-folding phantoms, where we applied a helical scan in a cone beam CT. The length of the phantoms did not significantly affect the dose distributions. The non-uniformity of the subtraction images between Figs. 8 (e) and (g) may be affected by the difference in the phantom length. Previous studies suggest that the diameter of the phantom, rather than its length, affects the dose.[31] Furthermore, the required length of a phantom for the helical scan and cone beam is currently under investigation.

Although XR-QA2 films were used in this study, GAFChromic EBT3 films also possess sensitivity in the dose range of 0.01–40 Gy, which includes the range of diagnostic X-ray energy. Few reports exist regarding X-ray CT dosimetry by the RCF method using GAFChromic EBT or EBT3 films.[16,32] It may be interesting to compare the dosimetry results of the XR-QA2 and EBT3 films. GAFChromic EBT3 film may also be a good candidate for dosimetry instruments. At first glance, our results appear to be different from those of previous studies.[23] Although previous studies using Monte Carlo simulation showed an increased dose in the centre of the phantom, our study shows the opposite. This phenomenon has already been explained by previous research. Gotanda et al. confirmed the centre dose was increased and the surface doses were decreased with decreasing phantom size.[33]

Our study has a few limitations. Our results are still preliminary. Our study was also performed for only one diameter; it is necessary to consider multiple diameters to envision various situations. Accurate quantitative assessment of dose distribution requires improvement in the accuracy of the calibration curve. In this study, we only present a qualitative distribution. However, this approach to X-ray CT dosimetry could become a useful method for dose distribution assessment with further study.

## V. CONCLUSION

In this study, the dose distributions with and without the presence of a CT phantom were investigated. Our results indicate that the dose distributions are higher without the phantom; however, no significant difference was observed in the doses with changes in the length of the phantom. To visualise the 2D dose distribution, 3D surface plots and subtraction images were calculated. In our research, multi-dimensional dose distribution analysis was useful. Longer phantoms may be useful for obtaining optimal doses in helical scans or cone beam CT.